# Generation of Long-Lived Isomeric States via Bremsstrahlung Irradiation

Yao Cheng[1,2], Bing Xia[1]
and
Chuanxiang Tang[1], Yinong Liu[1], Qingxiu Jin[1]

[1]*Department of Engineering Physics, Tsinghua University, Beijing*
[2]*Key Laboratory of Atomic and Molecular Nanosciences, Tsinghua University, Beijing*
yao@tsinghua.edu.cn

**Abstract**

A method to generate long-lived isomeric states effectively for Mössbauer applications is reported. We demonstrate that this method is better and easier to provide highly sensitive Mössbauer effect of long-lived isomers (>1ms) such as $^{103}$Rh. Excitation of (γ,γ) process by synchrotron radiation is painful due mainly to their limited linewidth. Instead, (γ,γ') process of bremsstrahlung excitation is applied to create these long-lived isomers. Isomers of $^{45}$Sc, $^{107}$Ag, $^{109}$Ag, and $^{103}$Rh have been generated from this method. Among them, $^{103}$Rh is the only one that we have obtained the gravitational effect at room temperature.

**Key words** long-lived Mössbauer effect, bremsstrahlung excitation

**1. Introduction**

We report the experimental observation of effective generation of Mössbauer emission by bremsstrahlung excitation and its particular Mössbauer effect. Since Mössbauer discovered the recoilless photon emitted from nuclei in 1958 [1], over a hundred isotopes have been investigated mainly using radioactive decay. Among them, $^{67}$Zn has been considered to be one of the most sensitive isotopes producing the Mössbauer effect [2]. The natural linewidth $\Gamma_0$ is defined by the lifetime $\tau_0$ according to the principle of uncertainty as:

$$\Gamma_0 = \frac{\hbar}{\tau_0} \tag{1}$$

We describe total linewidth $\Gamma$ including broadening $\Gamma_b$ of the natural linewidth as

$$\Gamma = \Gamma_0 + \Gamma_b, \tag{2}$$

which is assumed due to various impacts affecting the nuclear state. $\Gamma$ of $^{67}$Zn approached an ultra-fine level of 75 peV corresponding to a Q-value of $10^{15}$ to these 93 keV Mössbauer quanta [2]. Here the Q-value is

$$Q = \frac{E_\gamma}{\Gamma} \tag{3}$$

with corresponding gamma energy of $E_\gamma$. We arrange the stable Mössbauer isotopes in a sequence according to their $\Gamma_0$ as resolution from low to high, i.e., $^{73}$Ge, $^{181}$Ta, $^{67}$Zn, $^{45}$Sc, $^{109}$Ag, $^{107}$Ag, and $^{103}$Rh. The sequence remains the same by arranging them in terms of Q-value as sensitivity. The last four isotopes have isomeric lifetimes of the first excited states longer than millisecond. $^{103}$Rh studied in this report has $\Gamma_0$ of 0.1 aeV and $E_\gamma$ of 39.8 keV that correspond to a Q-value in the order of $10^{24}$.

Table 1: The features of the isomeric states of four isotopes.

| Isotope | $E_\gamma$(keV) | $\Gamma_0$(eV) | $t_{1/2}$(s) | a(%) | Estimated f(%) recoilless factor | | | $\alpha$ |
|---|---|---|---|---|---|---|---|---|
| | | | | | 300K | 77K | 0K | |
| $^{45}$Sc | 12.4 | $1.43\times10^{-15}$ | 0.318 | 100 | 77 | 90 | 93 | 400 |
| $^{107}$Ag | 93.1 | $1.03\times10^{-17}$ | 44.3 | 51.4 | $1\times10^{-5}$ | 0.1 | 4 | 20 |
| $^{109}$Ag | 88.0 | $1.15\times10^{-17}$ | 39.6 | 48.6 | $8\times10^{-5}$ | 0.4 | 6 | 20 |
| $^{103}$Rh | 39.8 | $1.35\times10^{-19}$ | 3366 | 100 | 45 | 70 | 74 | 1350 |

$E_\gamma$: Mössbauer transition energy, $\Gamma_0$: energy linewidth of the Mössbauer level, $t_{1/2}$: half life of the Mössbauer level, a: nature abundance, f: the Lamb-Mössbauer factor, $\alpha$: internal conversion coefficient.

## 2. Mössbauer spectroscopy with higher resolution

Higher sensitivity of Mössbauer effect beyond $^{67}$Zn is to be achieved by exploring the isomeric transition listed above, but almost no Mössbauer experiment applying these transitions have been carried out successfully. The dilemma predicament is twofold, i.e., poor sample features and poor experimental methodology. No reliable Mössbauer source is available to extend the experimental methodology from present limitations to further possibilities.

The low-lying isomeric state of $^{45}$Sc was discovered in 1964 [3]. It aroused interest worldwide because of its high Lamb-Mössbauer factor and its high sensitivity. Although $^{45}$Sc is superior to $^{67}$Zn in many aspects, it was never successfully applied due to many problems, i.e., poor efficiency of generating Mössbauer photon from β decay, inner bremsstrahlung induced from β decay, and high internal conversion coefficient [3]. By choosing isotopes for experiments, the measurement sensitivity and the ease of measuring process are two issues of particular concern. In this work, we report a method that successfully addresses these two issues. As considered at the beginning of this study, the sensitivity increase is meaningless, if new methodologies have not been developed. $^{45}$Sc shall be the first choice for us. However, our $^{45}$Sc excitation was successful several times but not repeatable. In the course of this study, we have recently found out the highly speed-up decay of $^{45}$Sc is probably due to the entangled Mössbauer emission. This report concentrates on the excitation of $^{103}$Rh.

The characteristics of interested long-lived isomers used in this work are listed in the Table 1. They have smaller $\Gamma_0$ than feV. We first noticed that the experimental condition was extremely tough for such a narrow linewidth. For example, it is difficult to perform the measurement with the corresponding Doppler-scan. The gravitational redshift significantly comes into play. Consequently, new measurement concepts are to be developed for this ultimate sensitivity. It seems that many extreme conditions such as vibration, homogeneity, and temperature *etc.* are necessary to preserve the natural linewidth. However, our observation on gravitational effect of $^{103}$Rh at room temperature leads us to think that many of these obstacles are not insuperable, which opens up new paths for Mössbauer spectroscopy. Generating such Mössbauer photon emission is just the beginning of the story. A good and easy-to-operate source becomes the major step for our future study.

**3. Excitation by bremsstrahlung**

A direct excitation with synchrotron radiation (SR) to generate Mössbauer photons with (γ,γ) process was suggested by Ruby in 1974 [4]. Ten years later, the excitation of $^{57}$Fe was carried out at the DORIS ring [5] and later became the well-known nuclear resonant scattering. Leupold *et al.* summarized the experimental results for many isotopes other than iron studied at the third generation synchrotrons such as ESRF, APS, and Spring 8 [6]. The most sensitive sample of their studies was $^{181}$Ta which was reported with less $\Gamma$ comparing to the radioactive samples prepared by ion bombardment. This observation is extremely important for the long-lived isomers; otherwise their natural linewidths would no longer contribute to the sensitivity. In any case, to increase the number of the photons in the excitation band is the baseline of an accurate measurement, but this flux increase is limited by $\Gamma_0$.

Ruby considered isotopes with $\Gamma_0$ ranging from $10^{-7}$ to $10^{-9}$ eV [4]. The long-lived isomers mentioned in the last paragraph have much smaller $\Gamma_0$. Is there any other effective method to achieve isomeric excitation? Coulomb excitation [7], (γ,γ') of the bremsstrahlung excitation [8-11], (γ,γ') of the photo-activation [12], and (n,γ) of the neutron excitation [13] are possible methods. Neutron excitation has the complication of generating radioactive isotope [13]. Two methods of (γ,γ') are considered to be more suitable for our application, since coulomb excitation generates more heat in samples. It is noted that $^{107}$Ag [8,9,10,12], $^{109}$Ag [8,10,12], $^{103}$Rh [11,12,13] created by (γ,γ') and (n,γ) have been reported, but not within the realm of the Mössbauer effect. $^{45}$Sc has been observed only by nuclear decay. An issue of whether these four isotopes might exhibit Mössbauer effect was next raised. A recent observation of gravitational redshift using $^{109}$Ag prepared from radioactive decay of $^{109}$Cd [14] provides positive encouragement. Alpatov *et al.* applied an experimental arrangement developed from the experience of decades of research. Surprisingly, their reported $\Gamma$ was only with the factor of 1 ~ 3 larger than $\Gamma_0$ [14]. We have understood the reason to obtain such a small broadening during the progress of this study, recently.

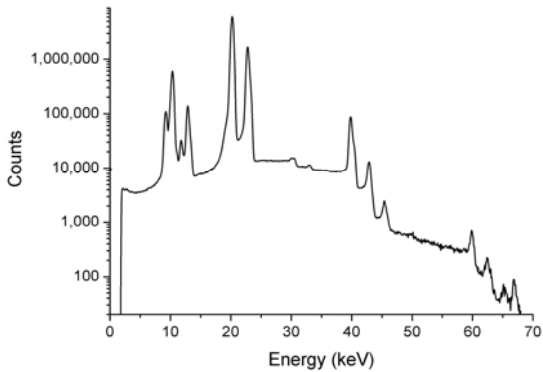

FIG. 1: Measured spectrum of the emissions from $^{103}$Rh isomers excited by bremsstrahlung irradiation from a 6 MeV linac for 120 min. The pile-up rejection system is off.

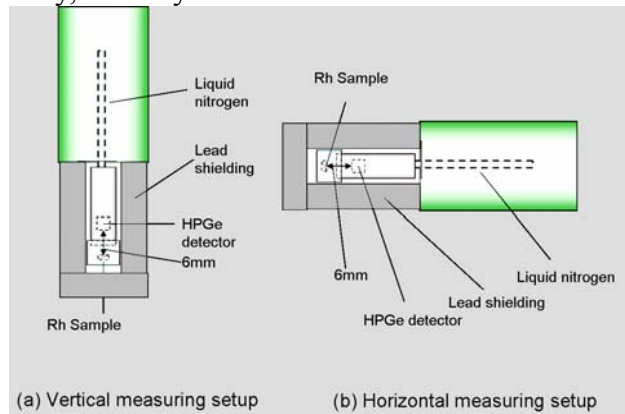

FIG. 2: (a) Vertical measuring setup. (b) Horizontal measuring setup. Configurations of rhodium sample and detector in the experiment of time-resolved spectroscopy. Setup (b) was directed in the E-W direction.

## 4. Experiment and results

We have carried out the isomeric excitation of $^{45}$Sc, $^{107}$Ag, $^{109}$Ag, and $^{103}$Rh using a 6 MeV standing-wave linac as the bremsstrahlung source [15]. The silver sample is a 999-purity φ-5cm coin of 3 mm in thickness and its natural abundance data are shown in Table 1. The scandium and rhodium samples (99.9% purity) purchased from Goodfellow are 0.1 mm-thick (1mm-thick) and have a 2.5×2.5 cm$^2$ area. The irradiation time on $^{103}$Rh was 120 min for the particular excitation illustrated in Figure 1. Three main emissions, i.e. 39.8, 20, and 23 keV are identified to be the E3 isomeric transition and its associated internal conversions. Figure 1 shows all of the possible pile-ups with the pile-up rejection system purposefully turned off. Two unknown peaks around 66 KeV are systematic [18]. The other two escape peaks of 10 keV and 13 keV are the features of our CANBERRA HPGe detector. Energy, pile-up time, dead time and the other system parameters are calibrated against $^{109}$Cd source.

## 5. Gravitational impact

$^{103}$Rh is chosen for our experiment of gravitational redshift due to its reasonable Lamb-Mössbauer factor at room temperature [15]. Here, we want to point out that the reported second order Doppler effect shall inhibit the resonant absorption [16]. Nevertheless, the Lamb-Mössbauer factor of rhodium at room temperature still exists from this extraordinary observation. The HPGe detector of CANBERRA BE3830 mounted on an in-situ object counting system is selected for our purpose, since it performs the freedom of setting any measurement angle corresponding to gravity (Fig. 2).

Figure 3 illustrates the measured K-shell X-rays and Mössbauer photons [15]. Data were accumulated every 2 minutes. The measured K X-rays in Fig. 3 behave as a nice exponential curve without significant dynamics. The fitted lifetime is 4856 s (Canberra DSP MCA, Inspector-2000, trapezoid shaping with 5.6μs/0.8μs as rise/flat-top time), almost the same as the tabled one [17]. However, Mössbauer photons decay faster (4712 ± 22 s). Speed-up decay of Mössbauer photons was found in series of experiment [15], by promptly rotating the detector from Fig. 2b to Fig. 2a. This enhanced decay is a function of excitation density as well as configuration corresponding to gravitation [15]. The anisotropic internal conversion rate demonstrates the anisotropic emission in two positions as shown in Figure 2. Through detail examination of the pile-up peaks in Figure 1, we found that the triple pile-up does not obey the statistic rule of photons [18]. Most of the observations including gravitational effect in the previous report [15] are related to this particular mechanism. The in-depth report on this new phenomenon is detailed in [18].

**Acknowledgements**

We thank Qiong Su and Jianping Cheng for the HPGe detector. Yao Cheng thanks Dr. Yaw-Wen Yang at NSRRC Taiwan for his carefully proofreading and Dr. Hognfei Wang at institute of Chemistry, Chinese Academy of Sciences for his intuitive advices. Yao Cheng thanks Professor Yuzheng Lin and his accelerator team in our department, for many efforts and his encouragement to study the reported issues. Yao Cheng thanks U. Van Bürck for fruitful discussions.

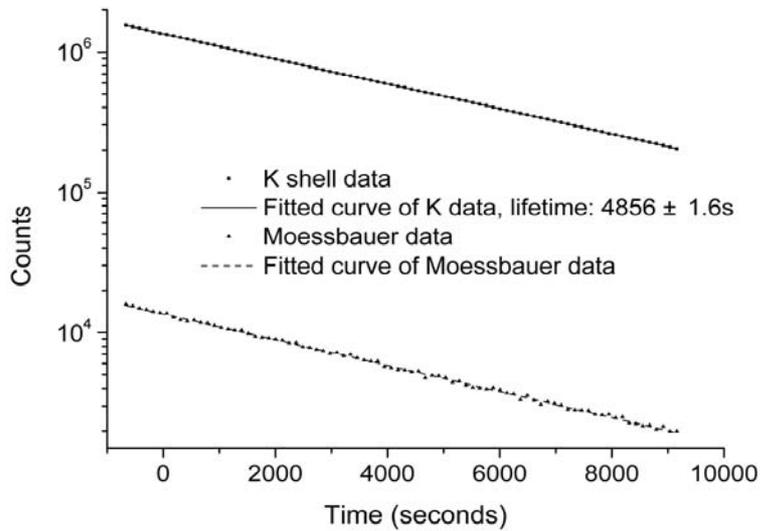

FIG. 3: K-shell and Mössbauer emissions measured in periods of two min. The detector was rotated from Fig. 2b to Fig. 2a at time zero. Mössbauer photon presents a speed-up decay.